\documentstyle[lprocldpf,epsf]{article}

\def\Journal#1#2#3#4{{#1} {\bf #2}, #3 (#4)}

\def\PRL{\em Phys. Rev. Lett.}
\def\PRB{{\em Phys. Rev.} B}

\def\p{{\bf p}}
\def\q{{\bf q}}
\def\om{\omega}

\begin{document}

\title{SHORT--RANGE ANTIFERROMAGNETIC CORRELATIONS\\
AND THE PHOTOEMISSION SPECTRUM}

\author{N. BULUT}

\address{Department of Physics, University of California,
Santa Barbara,\\ CA 93106--9530, USA}


\maketitle\abstracts{We present Quantum Monte Carlo results on the
antiferromagnetic correlations and the one--electron excitations of the
doped two--dimensional Hubbard model.
These results are helpful in interpreting the NMR, 
neutron scattering 
and photoemission experiments on the layered cuprates.
}
  
NMR and neutron scattering experiments have provided
clear evidence that the superconducting layered cuprates have 
short--range and low--frequency antiferromagnetic (AF) correlations.  
In addition, the photoemission experiments have shown that the 
one--electron excitations are heavily damped and strongly renormalized.
Perhaps, the simplest model to describe these properties is the 
two--dimensional Hubbard model on a square lattice,
\begin{equation}
H = -t\sum_{\langle i,j\rangle,\sigma} 
(c^{\dagger}_{i\sigma} c_{j\sigma} +
c^{\dagger}_{j\sigma} c_{i\sigma})
+U\sum_i c^{\dagger}_{i\downarrow} c_{i\downarrow}
c^{\dagger}_{i\uparrow} c_{i\uparrow},
\label{eq:ham}
\end{equation}
where $t$ is the hopping matrix element, $U$ is the onsite 
Coulomb repulsion, 
and $c_{i\sigma}$ annihilates an electron of spin $\sigma$ at site $i$.
Here, we present Quantum Monte Carlo (QMC) results on the 
magnetic fluctuations 
and the one--electron properties of the doped 2D Hubbard model.

\vspace{-0.5cm}
\begin{figure}[h]
\centerline{\epsfysize=7cm \epsffile[-30 184 544 598] {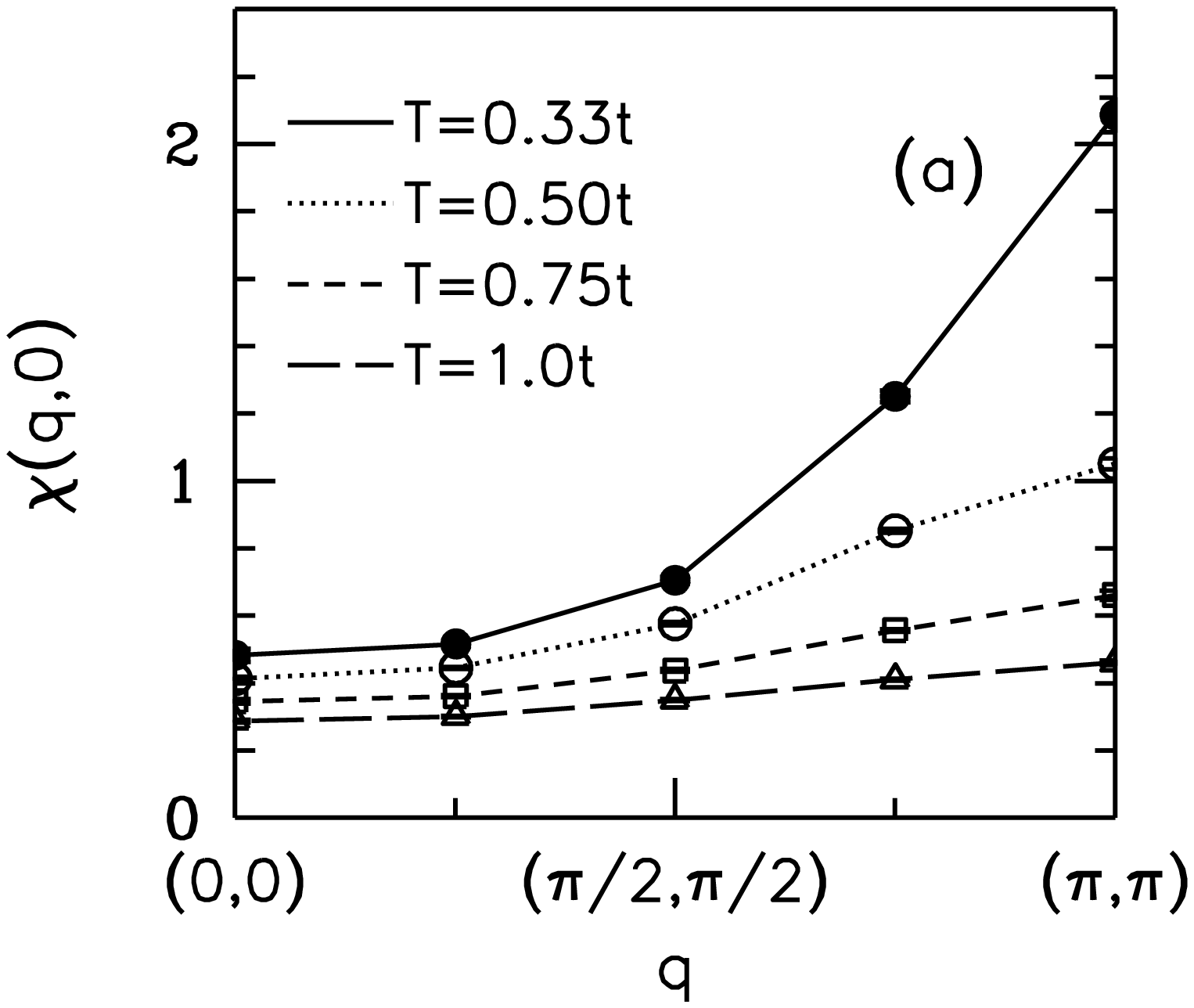}
\epsfysize=7cm \epsffile[98 184 672 598] {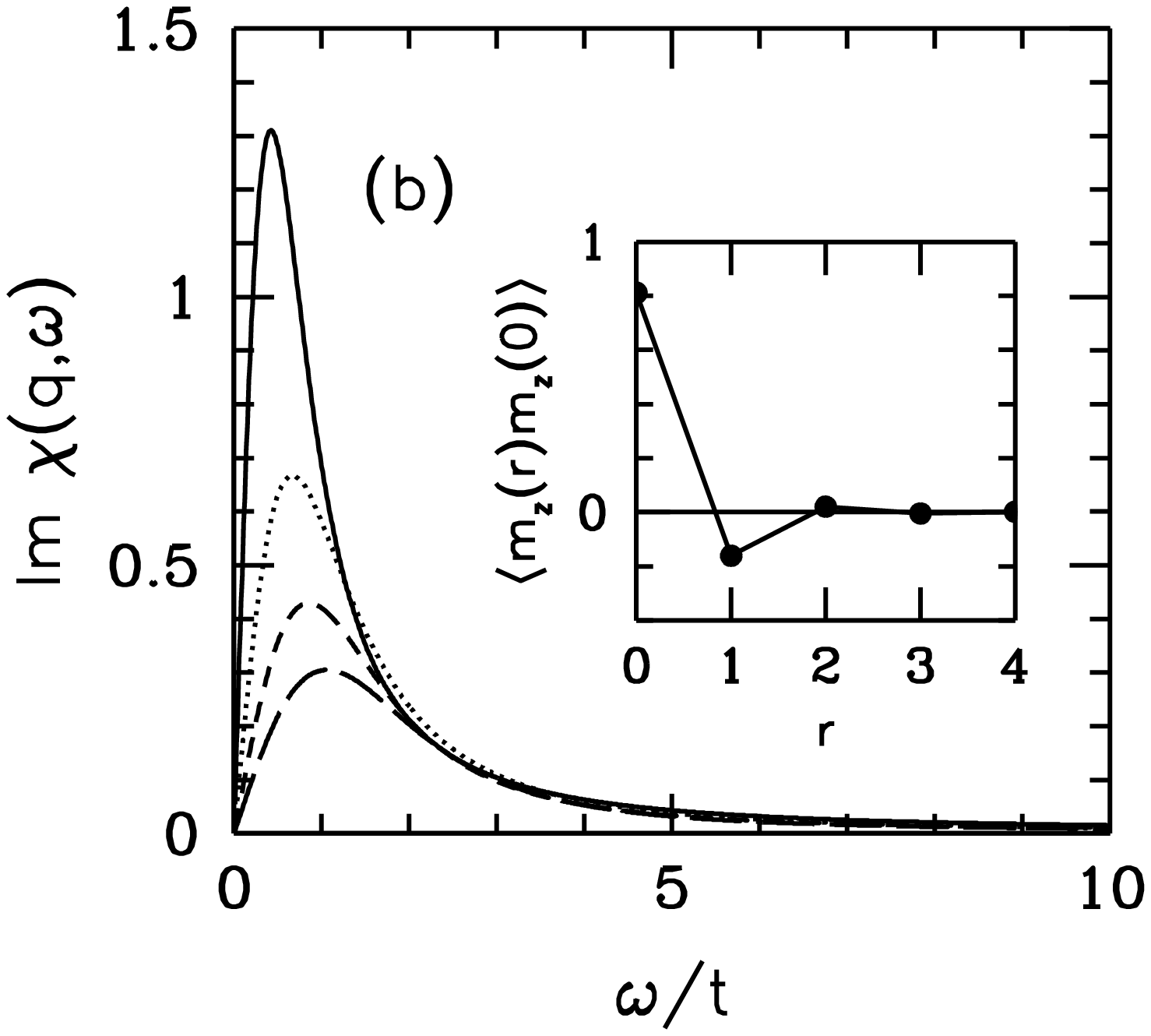}}
\caption{(a) Momentum dependence of the magnetic susceptibility 
$\chi(\q,\om=0)$ along the $(1,1)$ direction at various temperatures.
(b) Frequency dependence of the spin--fluctuation spectral weight 
${\rm Im}\,\chi(\q,\om)$ for $\q=(\pi,\pi)$ at the same temperatures
as in Fig. 1(a).
Inset: Real--space structure of the equal--time magnetization--magnetization
correlation function at $T=0.33t$.
\label{fig:fig1}}
\end{figure}

Using QMC simulations~\cite{QMC} and 
numerical analytic continuation methods~\cite{ME}
we have calculated the staggered magnetic susceptibility 
$\chi(\q,\om)$.
The results that we will present here are for $U=8t$ and 
$1/8$ doping on an $8\times 8$ lattice.
In addition, we use units such that $t=1$
and $\mu_B=1$.
In Figure 1(a), the momentum dependence of $\chi(\q,\om=0)$ 
is plotted along the $(1,1)$ direction. 
Figure 1(b) shows the spectral weight ${\rm Im}\,\chi(\q,\om)$ 
versus $\om$ for $\q=(\pi,\pi)$.
In the inset of Fig.~1(b), the equal--time
magnetization--magnetization correlation function is plotted as
a function of distance along the $(1,0)$ direction.
These figures show that there are short range AF correlations
in the 2D Hubbard model near half--filling.
NMR $T_1^{-1}$ and $T_2^{-1}$ measurements and magnetic neutron scattering
experiments have shown the existence of short--range AF 
correlations in the superconducting cuprates.

Next, we present results on the one--electron spectral weight 
$A(\p,\om) = - {\textstyle {1\over \pi}} {\rm Im}\,G(\p,\om)$,
where $G(\p,\om)$ is the one--electron Green's function.
$A(\p,\om)$ is of experimental interest, since,
within the sudden approximation, the photoemission intensity is 
proportional to $A(\p,\om)f(\om)$, where $f(\om)$ is the fermi factor.
In Figure 2, $A(\p,\om)$ versus $\om$ is plotted for 
three representative points of 
the Brillouin zone~\cite{BSW1,BSW2}.
Because of the many--body effects, 
the quasiparticle peak in $A(\p,\om)$ is damped,
and the quasiparticle bandwidth is reduced.
We also observe that $A(\p,\om)$ has significant 
dependence on temperature in this temperature regime.

In this article, we have presented a brief review of
the QMC data on the magnetic and 
one--electron excitations of the 2D Hubbard model near half--filling.
In this metallic regime, there exist low--frequency short--range
antiferromagnetic fluctuations, and the one--electron excitations
are heavily damped and strongly renormalized.
Similar electronic features have been observed experimentally for the
layered cuprates.

\begin{figure}[t]
\centerline{\epsfysize=5,1cm \epsffile[-207 184 367 598] {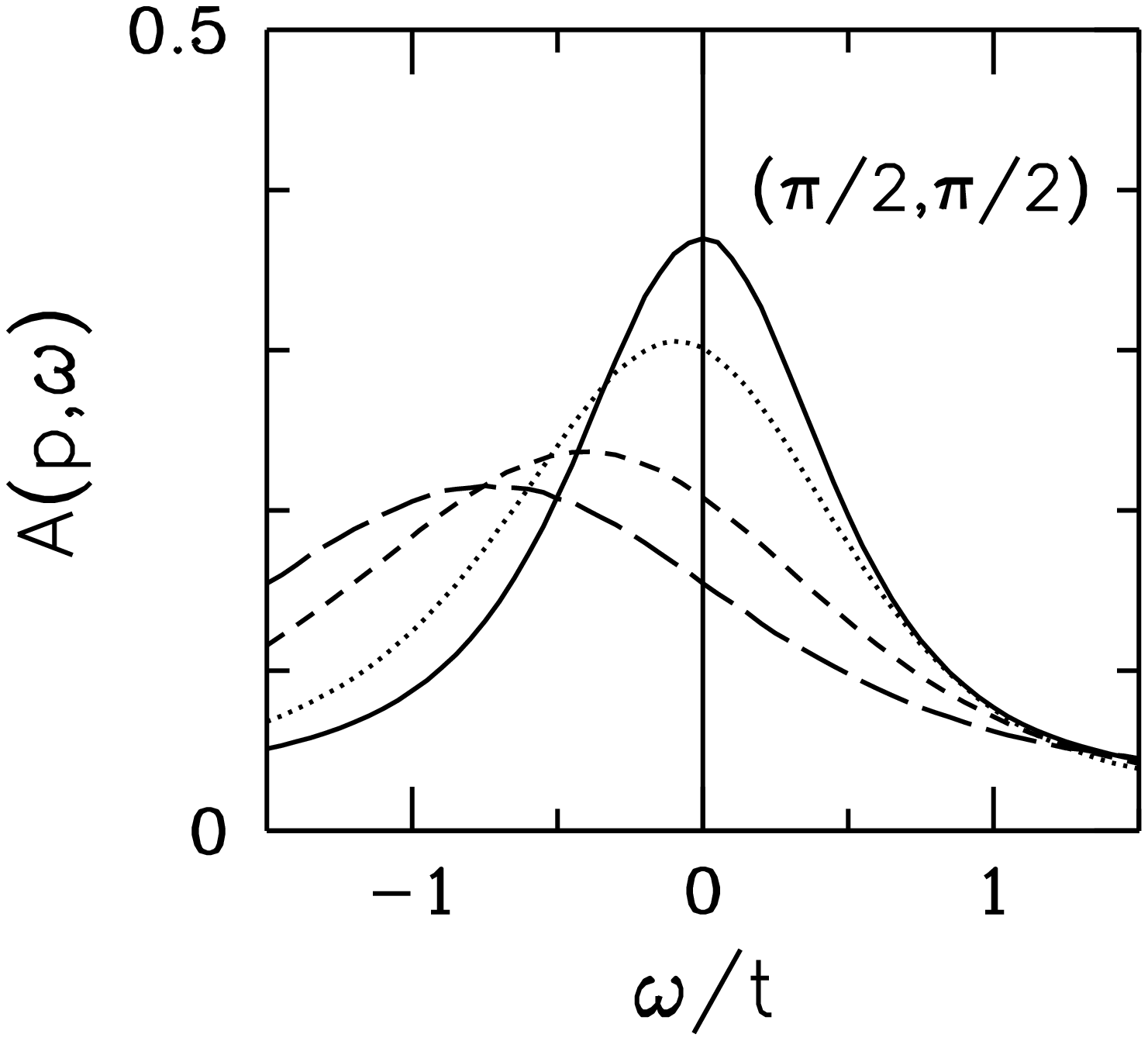}
\epsfysize=5.1cm \epsffile[18 184 592 598] {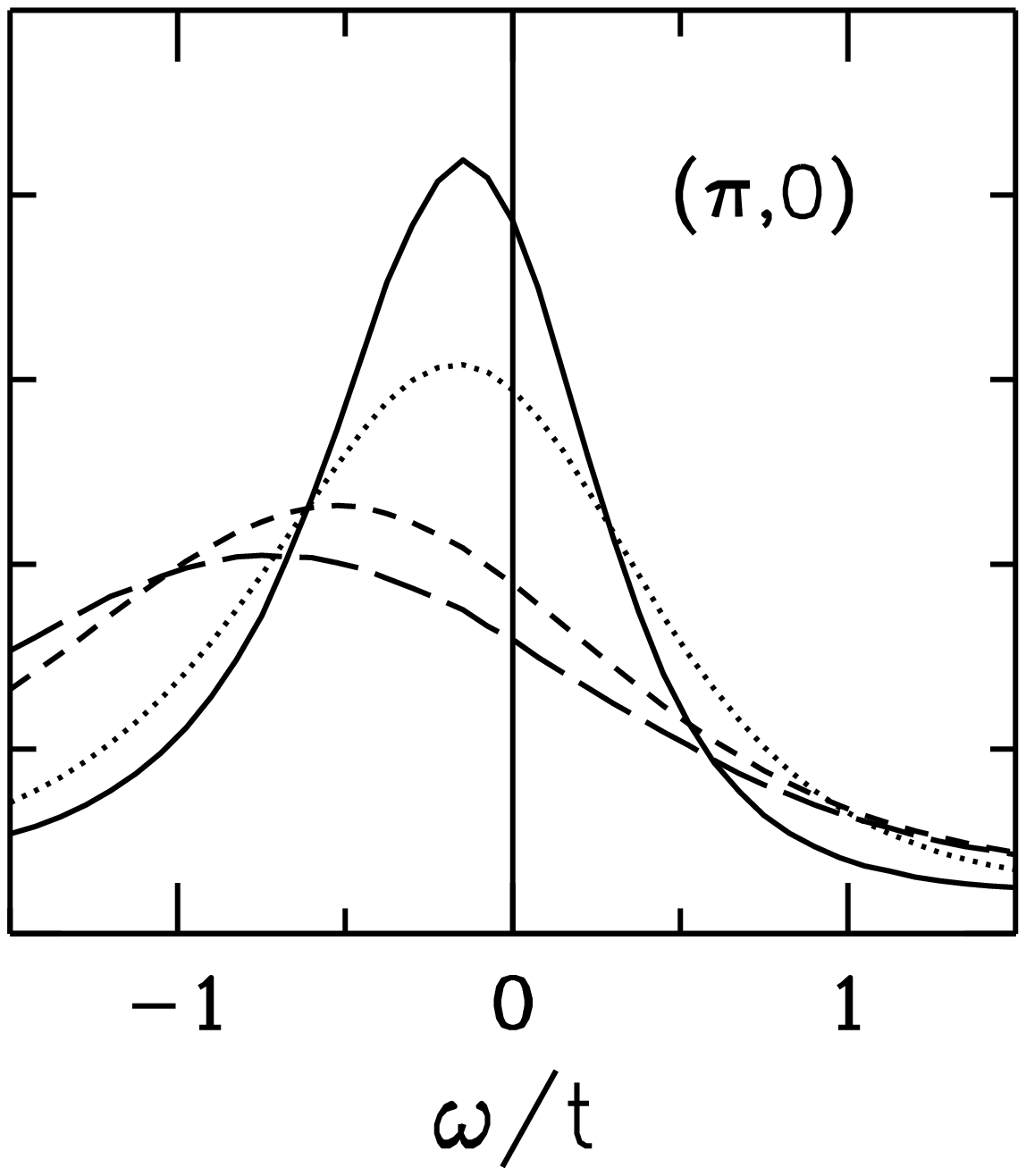}
\epsfysize=5.1cm \epsffile[243 184 817 598] {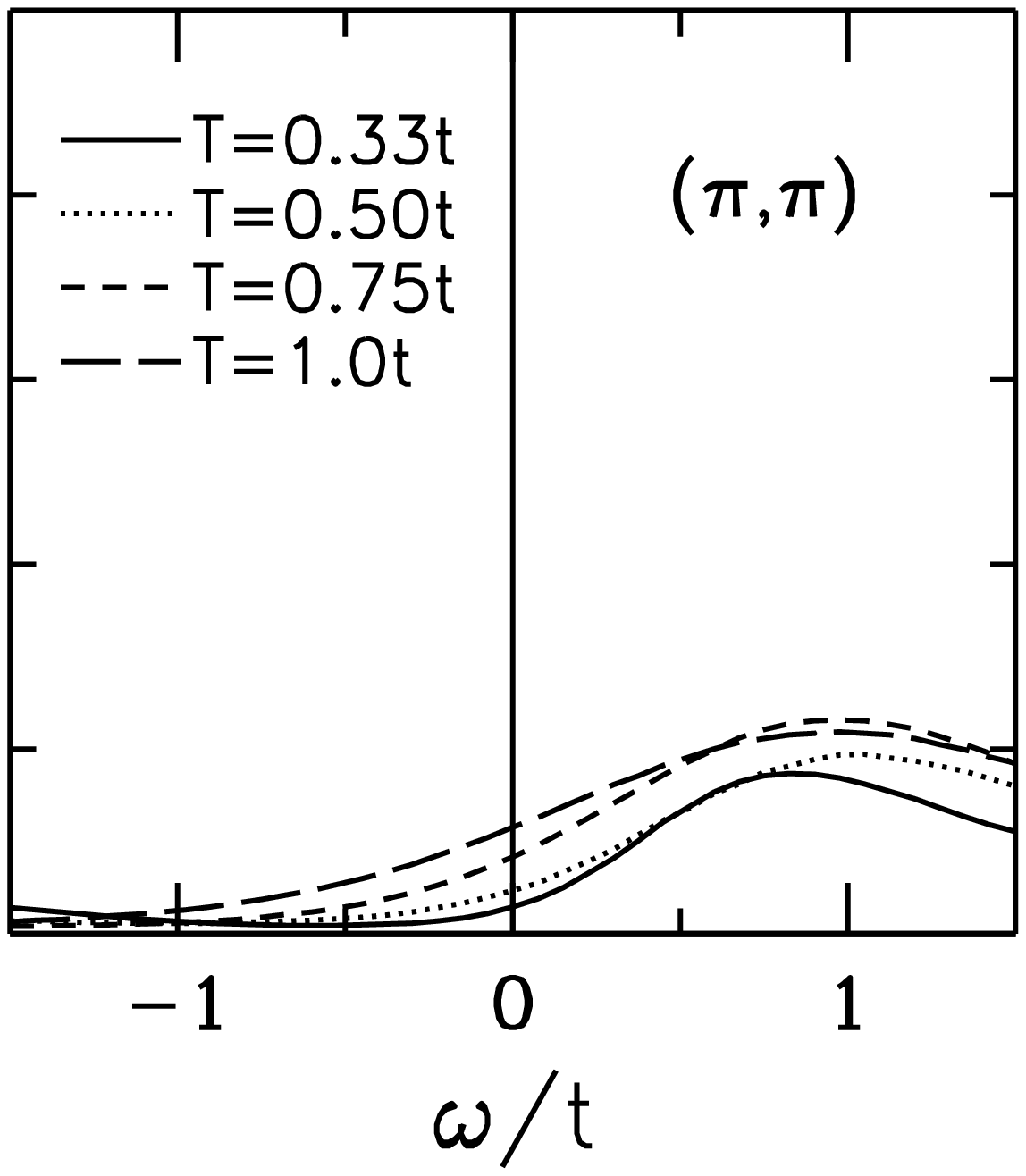}}
\caption{One--electron spectral weight $A(\p,\om)$ versus $\om$ 
at various temperatures for 
$\p=(\pi/2,\pi/2)$, $(\pi,0)$ and $(\pi,\pi)$.
\label{fig:fig2}}
\end{figure}

\section*{Acknowledgments}
The author would like to thank
D.J. Scalapino for many helpful discussions.
This review is based on work done in collaboration with 
D.J. Scalapino and S.R. White.
The author also gratefully acknowledges support from the 
National Science Foundation under Grant No. DMR92--25027.
The numerical calculations reported in this paper were performed 
at the San Diego Supercomputer Center.

\end{document}